# IN WIND TUNNEL THE REPRODUCTION OF VORTEX SHEDDING BEHIND CIRCULAR CYLINDERS AT HIGH REYNOLDS NUMBER REGIMES IS INCOMPLETE

Øyvind Mortveit Ellingsen
*LadHyX, CNRS-Ecole polytechnique, Palaiseau, France & CSTB, Nantes, France*

Xavier Amandolese
*LadHyX, CNRS-Ecole polytechnique, Palaiseau, France & LMSSC, CNAM, Paris, France*

Pascal Hémon
*LadHyX, CNRS-Ecole polytechnique, Palaiseau, France*

## ABSTRACT

*Wind tunnel tests of 2D rough cylinders are presented. The goal is to simulate the alternate vortex shedding in flow regimes encountered in wind engineering applications, where the full scale Reynolds number is larger than the one that can be reproduced in wind tunnel with small scaled models.*

*Measurements are mainly the synchronized unsteady wall pressures on the cylinder which are post processed using bi-orthogonal decompositions.*

*By comparing the small scale results with those from a previous large scale experiment, we show that the technique of rough cylinder is incomplete and can approach roughly global parameters only.*

## 1. INTRODUCTION

The circular cylinder is the bluff body which is one of the most studied bodies in aerodynamics along years. The circular shape induces indeed fundamental properties of flow, such as stall and unsteady wake and it has a great relevance in engineering applications. Especially in civil engineering there are numerous cases where circular cylinders of various diameters are submitted to wind and excitation by vortex shedding.

However the flow regime around this bluff body is extremely dependent on the Reynolds number which combines the effect of the cylinder's diameter $D$ and the mean wind velocity $\overline{U}$ such that:

$$Re = \frac{\overline{U}D}{\nu} \quad \#(1)$$

where $\nu$ is the air kinematic viscosity, as studied by (Adachi 1985; James et al. 1980; Warschauer & Leene 1971) and reviewed by Zdravkovich (1990).

The drag force coefficient of a 2D smooth circular cylinder is given in Figure 1 versus the Reynolds number, showing the data provided by the Eurocode (2005). For aerodynamic flows of practical applications with $Re$ greater than 10 000, three kinds of regime can be observed, namely subcritical with $Re \lesssim 200\,000$, critical if $200\,000 \lesssim Re \lesssim 600\,000$ and supercritical when $Re \gtrsim 600\,000$ (Roshko 1961; Lienhard 1966; Hoerner 1965; Simu & Scanlan 1978; Schewe 1983; Blevins 2001).

Following the same trend, the non dimensional frequency of the shedding, given by the Strouhal number

$$St = \frac{f\,D}{\overline{U}} \quad \#(2)$$

in which $f$ is the dimensional frequency, is plotted in Figure 2 (Shi et al. 1993; Adachi 1997; Zan 2008; van Hinsberg 2015; Ellingsen et al. 2022).

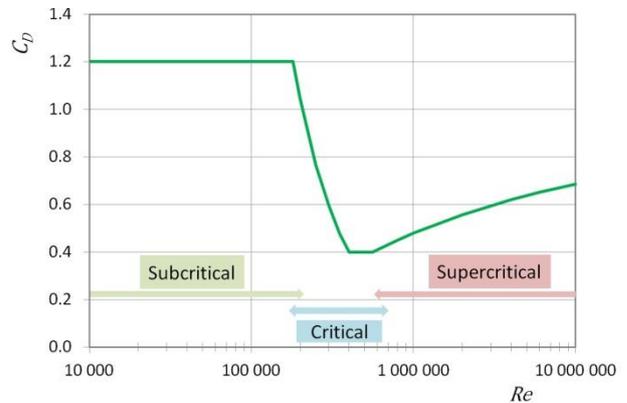

*Figure 1. $C_D$ versus $Re$ from Eurocode and definition of flow regimes*

In subcritical regime the boundary layer around the cylinder is laminar prior to its separation and the drag force coefficient $C_D = 1.2$. The alternate vortex shedding is well established and one observes that $St = 0.19 - 0.20$.

As the Reynolds number increases, the cylinder is subject to the "drag crisis" which characterizes the critical regime. This regime presents large variation of the drag force coefficient that decreases down to 0.4. The alternate vortex shedding is not well organized.

When the Reynolds number is further increased, reaching the supercritical regime, the drag coefficient $C_D$ is characterized by a smooth monotonic increase from 0.4 to 0.65. One can also observe a re-organization of the wake with an alternate vortex shedding having a Strouhal number subject to scattering, typically in the range 0.19 – 0.27. Recently in a large scale wind tunnel testing (Ellingsen et al 2022), we have shown that the scatter might be due to twin Strouhal numbers as shown Figure 2.

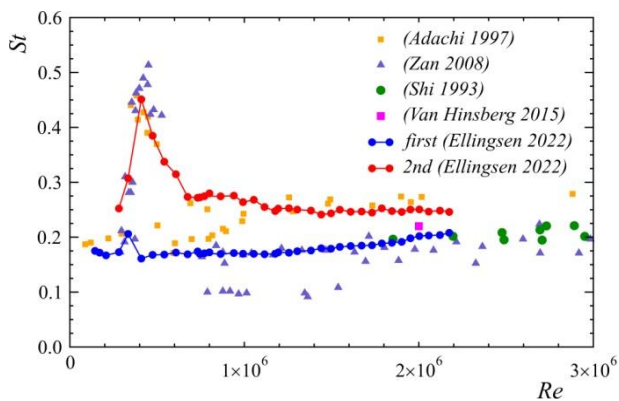

*Figure 2. St versus Re from different authors*

From these data, it turns out obvious that wind tunnel tests have to be made at the right Reynolds number which is encountered in the application. For instance in wind engineering (Lupi et al. 2017; Ellingsen et al. 2021), industrial chimneys have typically a diameter of 2 m and a natural first bending frequency of the order of 1 Hz. Then the critical wind velocity at which the resonance occurs with the alternate vortex shedding is in the range 7.5-10 m/s. The Reynolds number range for this case is then $10^6 – 1.3\ 10^6$, which is in the low region of the supercritical regime mentioned above.

But in practice for wind tunnel testing of such structures, scaled models typically of the order of 1/100 are used. This leads to a diameter of the chimney model of 2 cm and requires a wind velocity, in order to comply with the Reynolds number similarity, which is impossible to reach in a subsonic wind tunnel.

To compensate for this, a number of authors have considered the technique of added roughness on the cylinder model (Achenbach 1971; Szechenyi 1975, Achenbach & Heinecke 1981; Nakamura & Tomari 1982; Shih et al. 1993; Adachi 1997; van Hinsberg 2015). Rough cylinders are indeed known for shifting the drag crisis at smaller Reynolds numbers, depending on the roughness height. Global parameters such as the drag force coefficient $C_D$, the unsteady lift coefficient (RMS value) $C_\ell'$ and the Strouhal number $St$ are mainly used to calibrate added roughness techniques. But while it is used in wind tunnel testing (Barré & Barnaud 1995), the ability of such techniques to reproduce realistic supercritical flows is still being debated. The goal of this paper is to present wind tunnel results obtained with small scale cylinders equipped with roughness, in order to tentatively simulate the vortex shedding occurring at supercritical Reynolds number. Unsteady wall pressures are the main measured data and will be compared to those measured on a large scale cylinder from a previous experimental study (Ellingsen et al. 2022).

## 2. EXPERIMENTAL APPARATUS

The wind tunnel tests were performed in the NSA CSTB's wind tunnel in Nantes. The aerodynamic test section, 2x4 m², can reach a maximum wind speed of 30 m/s without turbulence generating grid and 19 m/s with it. The turbulence generating grid consists of semi-circular lengths of PVC tubing with the circular side towards the inflow and the flat side downstream. The grid components have a width of 0.05 m and the distance between two parallel lengths components are 0.2 m when measured from the center of each. When used, this grid was mounted 3.1 m upstream of the test model.

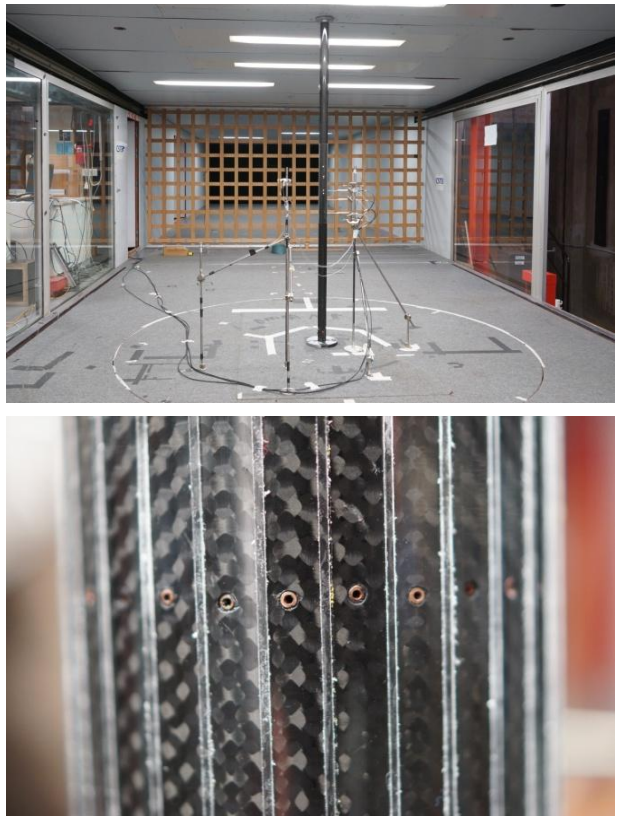

*Figure 3. Photo of the wind tunnel setup and detailed view of the pressure taps and ribs on the cylinder model*

A preliminary calibration in the empty test section was performed by mounting another Pitot tube at the cylinder model location to correlate the velocity seen by the model with the one measured by the reference Pitot tube. Cobra probes enabled characterization of the turbulence intensity in the three directions. For the configuration without turbulence generator, the turbulence is 1% in the main direction of the flow and 0.8% for the others. With the turbulence generating grid the turbulence intensity is 6% and 3% in the main direction and the two others respectively.

The test model consists of a circular cylinder made of carbon with diameter 0.055 m and vertically mounted in the wind tunnel, see Figure 3. It extends the entire 2 m height with the measurement location at mid-height. This section is equipped with 30 uniformly spaced pressure taps. The first pressure tap is placed at the stagnation point ($\theta = 0°$) and the rest spaced out uniformly with a separation of 12°. Vinyl tubing with length 1.0 m connects the pressure taps to a synchronized 32-channels pressure scanner (32HD ESP pressure scanner from Pressure Systems Inc.) with multiplex frequency of 70 kHz. The pressure scanner was rated up to 2500 Pa and have static errors within ± 0.03 %.

During the tests, the wind tunnel speed is kept constant for each measurement point and all the measured signals are recorded during 180 s at the sampling frequency of 400 Hz.

The cylinder has a first damped natural frequency of 25 Hz and a critical damping ratio of 2.1% which make the vibration level negligible.

To add roughness to the circular cylinder, ribs with rectangular cross-section at 12° intervals are attached on the circumference. Ribs having a cross section of constant width 0.8 mm and three different thicknesses $h =0.2$, 0.5 and 1 mm are tested. They are made with acrylic sheets of different thicknesses from which ribs are obtained with a numerically controlled laser cutter. The resulting non-dimensional roughness are $k = h/D = 3.6 \cdot 10^{-3}$, $9.1 \cdot 10^{-3}$ and $18.2 \cdot 10^{-3}$.

## 3. GLOBAL RESULTS

The drag coefficients and the unsteady lift coefficients are shown in Figure 4 for the smooth flow and turbulent flow conditions.
One can observe that each roughness configuration promotes an earlier critical regime, down to Reynolds number values at least ten times lower than for the reference values reported in Figure 1. However, the supercritical regime is not fully reached for the thinner ribs with a thickness 0.2 mm. This is in agreement with the assumption based on Szechenyi's results (Szechenyi 1975) that the supercritical regime cannot be reached for $Re_h < 200$, (the Reynolds number based on the rugosity height $h$) which is the case here for the smaller roughness when $Re < 55\,000$.

Globally the added upstream turbulence (from 1 to 6 %) promotes the critical regime to occur earlier but does not change drastically the flow nature. However the roughness height is of major importance in the phenomenon. For the roughness, 0.5 mm, the supercritical regime seems to be reached for $Re > 40\,000$ and even earlier than the minimum wind tunnel speed for the roughness height of 1 mm.

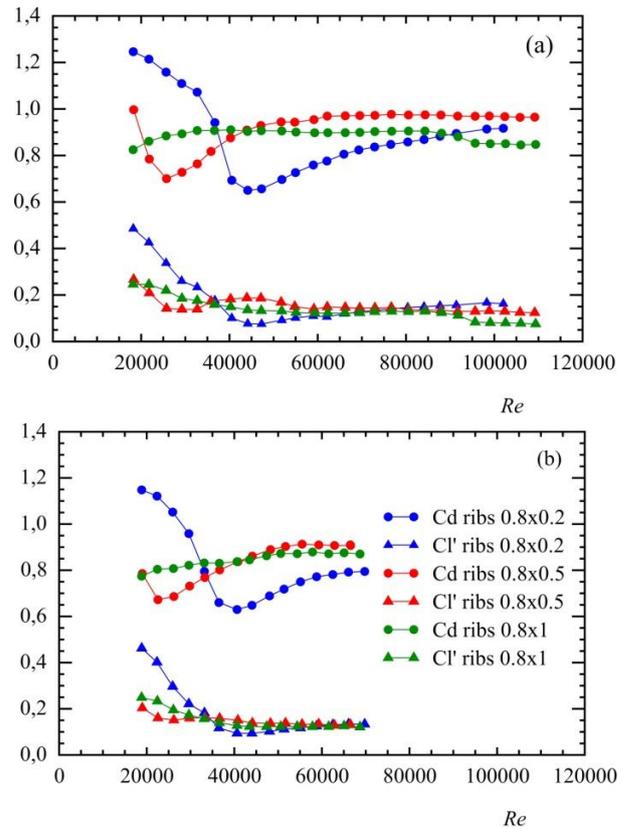

Figure 4. $C_D$ and $C_\ell'$ versus $Re$ for rough cylinders; (a) smooth flow; (b) turbulent flow

The Strouhal number is obtained with the PSD of the lateral velocity component issued from the Cobra probes. Main results are given in Figure 5 with few data points from (Adachi 1997) obtained with a cylinder equipped with roughness $k = 2.54 \cdot 10^{-3}$ which is a little lower than the smallest ribs used here. Globally there is agreement although experimental conditions are not exactly similar.

For the medium size roughness 0.5 mm, the differences between smooth and turbulent flows follow what has been previously observed on the drag force. Beyond $Re = 40\,000$ the Strouhal number doesn't change and remains in the range [0.21-0.22].

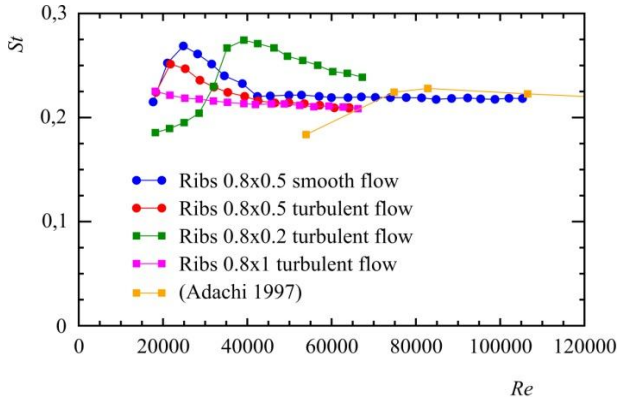

*Figure 5. St versus Re for rough cylinders.*

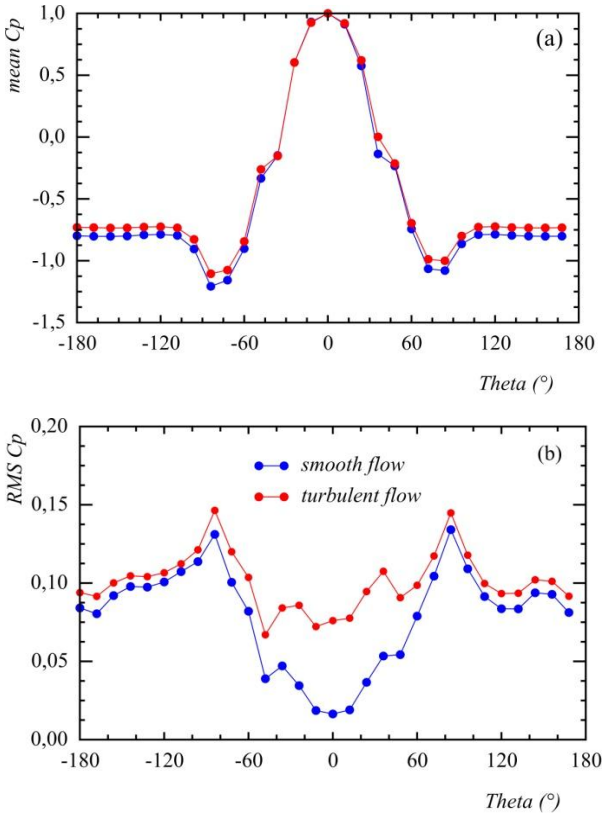

*Figure 6. Pressure coefficients for ribs 0.8x0.5 at Re=66 000; (a) time averaged; (b) RMS.*

The choice of the "best" roughness height to reproduce the supercritical regime is not trivial, but might be done with simple criteria. Firstly we must note that the evolution of the drag coefficient with Reynolds number and that of the Strouhal number are reproduced with the intermediate value 0.8x0.5 mm: in the low Reynolds number range, the critical region is visible and the stabilization of the coefficients is reached further. On the contrary, the small roughness height does not show any stabilization, and the highest roughness has no critical regime. This is of course in relation with the capacities of the wind tunnel to provide a sufficient wide range of velocities, in smooth and turbulent flow.

From that single argument we may assume that the roughness 0.8x0.5 mm must be selected. A deeper analysis is performed in the next section with the analysis of the wall pressure distributions and further comparisons are given in the conclusion.

## 4. PRESSURE DISTRIBUTION

The wall pressure distribution is shown in Figure 6 for the ribs 0.8x0.5 mm at the Reynolds number 66 000 for smooth and turbulent flow. The pressure coefficient is defined as

$$Cp(\theta, t) = \frac{P(\theta, t) - P_{ref}}{\frac{1}{2}\rho \bar{U}^2} \quad \#(3)$$

where $P(\theta, t)$ is the instantaneous measured pressure at the azimuth angle $\theta$. The reference pressure $P_{ref}$ is the mean static pressure in the wind tunnel obtained from the reference Pitot tube and $\rho$ is the air density corrected by atmospheric pressure and air temperature.

The differences between the two upstream flow cases are small, except for the standard deviations in the front region of the cylinder section ($-45° < \theta < 45°$) where the turbulent flow contributes to an increase of the fluctuations.

## 5. ANALYSIS OF THE WALL PRESSURE

In this section we use the bi-orthogonal decomposition (BOD) of the wall pressure signals in order to better analyze the results.

### 5.1. The bi-orthogonal decomposition

We recall here the analyzing technique which was first introduced by (Aubry et al. 1991). The idea of the BOD is to decompose the spatio-temporal signal $Cp(\theta, t)$ in a series of spatial functions $\phi_i(\theta)$ named further as "topos", coupled with a series of temporal functions $\psi_i(t)$ named "chronos". The BOD can be written as

$$Cp(\theta, t) = \sum_{i=1}^{N} \alpha_i\, \phi_i(\theta)\, \psi_i(t) \quad \#(4)$$

where $\alpha_i$ are the eigenvalues of the spatial or the temporal covariance matrix of the signal $Cp(\theta, t)$. $N$ is the number of terms retained for the decomposition. Chronos and topos are orthogonal between them and normed. Mathematical details can be found in (Aubry & Lima 1991) and practical applications are presented in (Hémon & Santi 2003).

It was shown that the eigenvalues $\alpha_i$ are common to chronos and topos and that the series con-verge rapidly so that $N$ is possibly small compared to the original size $T$ of the problem (the smallest between the number of pressure taps and the number of time records). This means that the $\alpha_i$ have a numerical value that decreases rapidly. Their sum $A = \sum_{i=1}^{T} \alpha_i$ represents the total energy in the original signal. Then each couple of chronos and topos have their contribution to the signal which decreases as long as their rank $i$ increases.

Moreover, by proper spatial integration of the topos $\phi_i(\theta)$ it is possible to determine the contribution of that term to the drag or the lift force, having in mind the corresponding chronos $\psi_i(t)$ which provides the time evolution of the couple $\phi_i(\theta)\,\psi_i(t)$. Hence the PSD of chronos can also be calculated in the same way than the cobra probes signals in order to determine their main frequency content.

Note that BOD is very similar to proper orthogonal decomposition (POD), except that the mean value of the original signal is kept in the analysis, refer to (Hémon & Santi 2003) for a discussion on that point.

### 5.2. BOD of wall pressure signals

The analysis is performed with the cylinder equipped with the roughness 0.8x0.5 mm which is assumed to be the best configuration for approaching the supercritical regime.

It appears that the BOD turns out very efficient in compressing the data with the first two terms taking almost 99.9 % of the energy. The first one corresponds to the mean value of the pressure distribution, producing the static drag, while the second one is the main unsteady component, producing the unsteady lift generated by the vortex shedding.

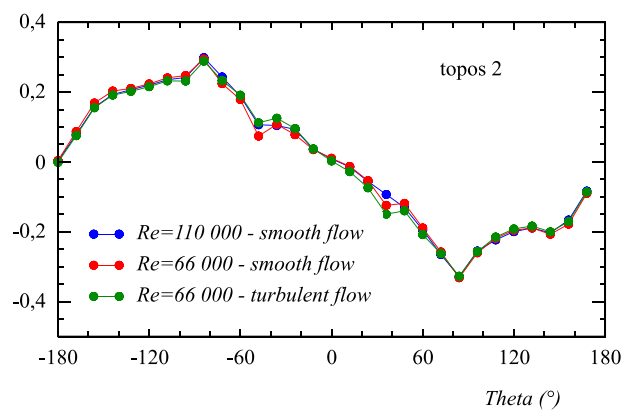

*Figure 7. Topos 2 for ribs 0.8x0.5 in different flow conditions.*

The topos 2 is shown in Figure 7 for different flow conditions. We note that, in the range explored, the shape of the topos 2 is not influenced by the Reynolds number or the flow turbulence. It is characterized by two anti-symmetrical bumps with their maximum located at $\theta = \pm 80°$, which creates a strong unsteady lift.

The topos of higher rank are more noisy and do not present a clear spatial organization which could be linked to the alternate vortex shedding phenomenon.

This is coherent with the lack of emerging frequency in the corresponding chronos, while the Strouhal number found with the PSD of the chronos 2 is in full agreement with the one which was found in the wake (see Figure 5), either in smooth or turbulent flow.

Determining the contribution of the topos 2 to the unsteady lift versus Reynolds number, we find an almost constant contribution of 87-90 % of the total force. Therefore the second term of the BOD is clearly the term which generates the lift force due to the alternate vortex shedding.

## 6. COMPARISON WITH THE SUPERCRITICAL REGIME

The comparison of the results obtained with the rough cylinder is performed with the results of Ellingsen et al. (2022) where the same kind of measurements were performed with a large smooth cylinder in a large wind tunnel. The Reynolds number in supercritical regime was achieved up to 2 170 000.

A first comparison on global parameters can be done using Tables 1 and 2. Few points must be pointed out:
- The drag coefficient of the rough cylinder is larger (+ 76 % than for the supercritical flow)
- The unsteady lift is larger also (+ 15 %)
- There is only one Strouhal number which value is between the twin Strouhal numbers observed in the true supercritical regime.

| | | | |
|---|---|---|---|
| $C_D$ | 0.97 | $C_\ell'$ | 0.146 |
| $St$ | 0.22 | | |
| $Cp_{min}$ | -1.2 | Location $\theta$ | $\pm\,80°$ |
| $Cp_{max}'$ | 0.13 | Location $\theta$ | $\pm\,80°$ |

*Table 1. Main results for ribs 0.8x0.5 at Re=66 000 in smooth flow*

| | | | |
|---|---|---|---|
| $C_D$ | 0.55 | $C_\ell'$ | 0.127 |
| $St_1$ | 0.20 | $St_2$ | 0.25 |
| $Cp_{min}$ | -2.5 | Location $\theta$ | $\pm\,80°$ |
| $Cp_{max}'$ | 0.3 | Location $\theta$ | $\pm\,110°$ |

*Table 2. Main results for a smooth cylinder at Re=2 000 000 (Ellingsen et al. 2022)*

There are also significant differences in the pressure distribution, as it can be seen in the Figure 8, for both the time averaged and RMS values. In particular the minimum, while located at the same place, is almost twice (negatively) for the reference supercritical configuration. The RMS distribution exhibits even greater differences, having its maximum at different places, $\pm 80°$ for the rough cylinder against $\pm 110°$ for supercritical flow. Moreover the maximum RMS values are more than doubled in supercritical flow (0.3 against 0.13).

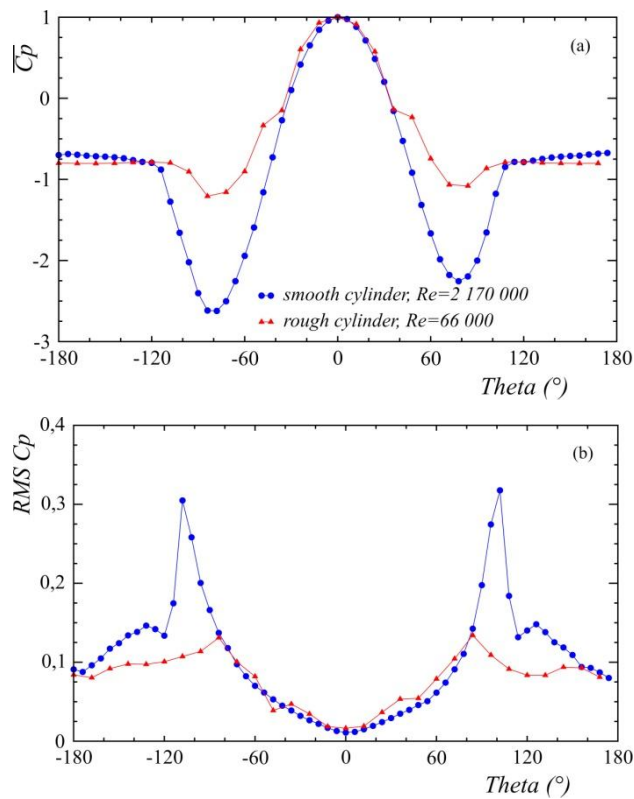

Figure 8. Comparison of the pressure distribution of the cylinder with ribs 0.8x0.5 at Re=66 000 in smooth flow and a smooth cylinder at Re=2 170 000 from (Ellingsen et al. 2022) (a) time averaged (b) RMS value

The tendencies observed on the RMS pressure distribution are found in the analysis through the BOD. The topos 2 in the two cases is shown in Figure 9. This term is responsible of 90 % of the total unsteady lift at a single Strouhal frequency for the rough cylinder.

In true supercritical regime, the same term represents also 89 % of the unsteady lift but at the first Strouhal frequency. Another BOD term (number 4, see (Ellingsen et al. 2022)) associated to a second Strouhal frequency is necessary to recover the unsteady lift force.

The unsteady pressure distribution which generates the unsteady lift force due to vortex shedding is therefore very different in the two cases.

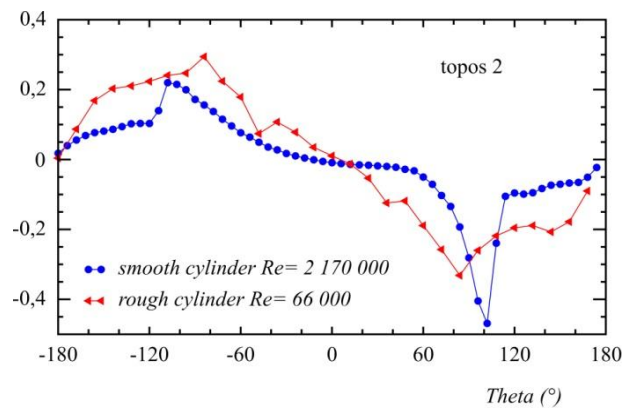

Figure 9. Comparison of the topos 2 of the cylinder with ribs 0.8x0.5 at Re=66 000 in smooth flow and a smooth cylinder at Re=2 170 000 from (Ellingsen et al. 2022).

## 7. CONCLUSIONS

Small scale experiments with a cylinder equipped with artificial roughness show that the flow at supercritical regime can be roughly approached, as on the unsteady lift coefficient, but it remains far from being reproduced. The second Strouhal number is absent and the measured value is between the two ones of the supercritical regime. The corresponding unsteady lift is only recovered by the second structure in the BOD. Moreover the shapes of the topos are quite different: the lift production at supercritical Reynolds number is concentrated in a narrow region of the azimuth angle, while the lift is produced in a wider range for the artificially simulated supercritical flow case.

## 8. ACKNOWLEDGEMENTS

This work is part of a partnership co-funded by Beirens (Poujoulat Group), Centre Scientifique et Technique du Bâtiment (CSTB), Centre National d'Etudes Spatiales (CNES) and LadHyX, CNRS-Ecole polytechnique.

Special acknowledgement is extended to Olivier Flamand from CSTB for the wind tunnel operation.